\documentclass[aps,prl,twocolumn,showpacs,preprintnumbers,nofootinbib,superscriptaddress]{revtex4}
\usepackage{graphics}  
\usepackage{epsfig}
 \usepackage{verbatim}

\IfFileExists{srcltx.sty}{\usepackage[active]{srcltx}}

\newcommand{\earth}{\ensuremath{\oplus}}

\newcommand{\be}{\begin{equation}}
\newcommand{\ee}{\end{equation}}
\newcommand{\bea}{\begin{eqnarray}}
\newcommand{\eea}{\end{eqnarray}}

\begin{document}

\title{Neutrinos from the terrestrial passage of supersymmetric dark-matter Q-balls}

\author{Alexander Kusenko}
\affiliation{Department of Physics and Astronomy, University of California, Los
Angeles, CA 90095-1547, USA}
\affiliation{Institute for the Physics and Mathematics of the Universe,
University of Tokyo, Kashiwa, Chiba 277-8568, Japan}
\author{Ian M. Shoemaker}
\affiliation{Department of Physics and Astronomy, University of California, Los
Angeles, CA 90095-1547, USA}

\preprint{UCLA/09/TEP/50}

\begin{abstract}
Supersymmetry implies that stable non-topological solitons, Q-balls, could form in the early universe and could make up all or part of  dark matter.   We show that the relic Q-balls passing through Earth can produce a detectable neutrino flux.  The peculiar zenith angle dependence and a small annual modulation of this flux can be used as signatures of dark-matter Q-balls. 
\end{abstract}

\maketitle

All phenomenologically acceptable supersymmetric extensions of the standard model predict the existence of  Q-balls~\cite{Kusenko:1997zq}.   Some of these Q-balls are stable~\cite{Dvali:1997qv}; they could form in the early universe, and they can now exist as a form of dark matter~\cite{Kusenko:1997si}. The cosmology of supersymmetric Q-balls has been studied extensively~\cite{Frieman:1988ut,Frieman:1989bx,Griest:1989bq,Griest:1989cb,Enqvist:1997si,
Enqvist:1998ds,Enqvist:1998xd,Kusenko:1997ad,Kusenko:1997hj,Kusenko:1997it,Kusenko:1997vp,Laine:1998rg,Enqvist:1998en,
Enqvist:1998pf,Axenides:1999hs,Banerjee:2000mb,Battye:2000qj,Allahverdi:2002vy,Enqvist:2003gh,Dine:2003ax,Kusenko:2004yw,
Kusenko:2005du,Berkooz:2005rn, Berkooz:2005sf,Kusenko:2008zm,
Johnson:2008se,Kasuya:2008xp,Sakai:2007ft,Campanelli:2007um,Kasuya:2000wx,
Kawasaki:2005xc,Kasuya:2007cy,Shoemaker:2008gs,Campanelli:2009su}.  

The search for relic Q-balls so far has been focused on their ability to change the baryon and lepton number of fermions scattering off a  Q-ball~\cite{Dvali:1997qv,Kusenko:1997vp,Kusenko:2004yw}.  A relic Q-ball passing through matter can interact with electrons, protons, and neutrons, converting them into their respective antiparticles with probability of the order of one~\cite{Kusenko:2004yw}.  The change in the baryon and lepton numbers of the matter fermions is compensated by the change in the baryon and lepton number of the Q-ball.  This process can produce a trail of $p\bar{p}$ and $n\bar{n}$ annihilations following the passage of a relic Q-ball through  matter.  This property has been used to set astrophysical limits~\cite{Kusenko:1997it,Kusenko:2005du}, as well as the limits based on the non-observation of Q-ball passages through Super-Kamiokande and other detectors~\cite{Kusenko:1997vp,Arafune:2000yv,Takenaga:2006nr}.  We will present a new signature of the relic Q-balls: neutrinos produced from the passage of Q-balls in Earth can give an observable signal in neutrino detectors. 

We calculate the flux and the spectrum of neutrinos produced in the $p\bar{p}$ and $n\bar{n}$ annihilations accompanying the passage of the relic Q-balls through Earth.  Such neutrinos should be observable in neutrino detectors.  The peculiar zenith dependence of the neutrino flux can help distinguish the flux due to Q-balls from that due to atmospheric interactions of cosmic rays.  In some range of parameters, the neutrinos may provide a better sensitivity to the relic Q-balls than the search strategy based on  direct detection~\cite{Kusenko:1997vp,Arafune:2000yv,Takenaga:2006nr}.

The primordial generation of stable Q-balls requires the formation of an Affleck-Dine scalar condensate\cite{Affleck:1984fy,Enqvist:2003gh,Dine:2003ax} which forms along one of the many flat directions of the MSSM.  
The effects of supersymmetry breaking determine the potential along the flat directions, which we will assume to be 
\be V(\phi) =M_{s}^{4} \log \left(1 + \frac{|\phi|^{2}}{M_{s}^{2}}\right),  \ee
as in the case of gauge-mediated supersymmetry breaking scenario. The scale $M_{s}$ is determined by supersymmetry breaking. 
The Q-ball radius $R(Q_B)$ and mass $M(Q_B)$ in this potential are described by the following relations~\cite{Dvali:1997qv}: 
\be 
R(Q_B) \approx M_s^{-1} Q_B^{1/4}, \  M (Q_B) \approx M_s Q_B^{3/4} 
\label{mass_rels}
\ee
The Q-ball is rendered stable against decay into nucleons as long as the mass of the scalar baryons is less than a nucleon mass: $\omega=M(Q_B)/Q_B < 1\, {\rm GeV}$, which corresponds to $Q_B \gtrsim 10^{12} \left(M_{s}/ {\rm TeV}\right)^{4}$.  

If Q-balls make up all of the cosmological dark matter, the number of Q-balls found inside Earth at any given time is 

\be N_{\earth} \sim \frac{\rho_{_{DM}} V_{\earth}  }{M(Q_B)}  \sim 3 \times 10^{5}
\left( \frac{10^{24}}{Q_B} \right)^{3/4} \left( \frac{{\rm TeV}}{M_{s}} \right), \ee
where $\rho_{_{DM}}  \sim 0.3\, {\rm GeV}/{\rm cm}^3$ is the local dark matter density, and $V_{\earth} $ is the Earth's volume. Since the existing experimental limits \cite{Kusenko:1997vp,Arafune:2000yv,Takenaga:2006nr} imply $Q_B \gtrsim 10^{24}$, and since cosmological scenarios are consistent with such large Q-balls, we choose $Q_B\sim 10^{24}$ as a representative reference point.  The number of neutrinos generated by the Q-balls passing through Earth is  
\bea \frac{dN_{\nu}}{dt} &\sim& 10~ (\pi [R(Q_B)]^2)~ n_{n}~ v_{0}~ N_{\earth} \nonumber \\
&\sim&  5\times10^{17} \left( \frac{10^{24}}{Q_B} \right)^{1/4}
\left(\frac{{\rm TeV}}{M_{s}}\right)^{3} {\rm s}^{-1}, \eea
where $n_n$ is the average nucleon density, $v_{0} \sim 10^{-3}$ is the virial velocity of the dark matter particles, and we have taken it into account that $\sim 10$ neutrinos are produced in every $p \bar{p}$ annihilation.  

The flux at the Earth's surface is, therefore, 
\be F_{\nu, \earth} \sim \frac{1}{4 \pi R_{\earth}^{2}} \frac{dN_{\nu}}{dt} \sim 0.1 \left( \frac{10^{24}}{Q_B} \right)^{1/4}
\left( \frac{{\rm TeV}}{M_{s}}\right)^{3} {\rm cm}^{-2} {\rm s}^{-1} 
\label{terrestrial_flux}
\ee
Although more neutrinos are produced by Q-balls going through the interior of Sun, the flux on Earth is dominated by terrestrial neutrinos because of the $1/r^{2}$ suppression for the solar flux. 

\begin{figure}
\centering
\begin{tabular}{cc}
\epsfig{file=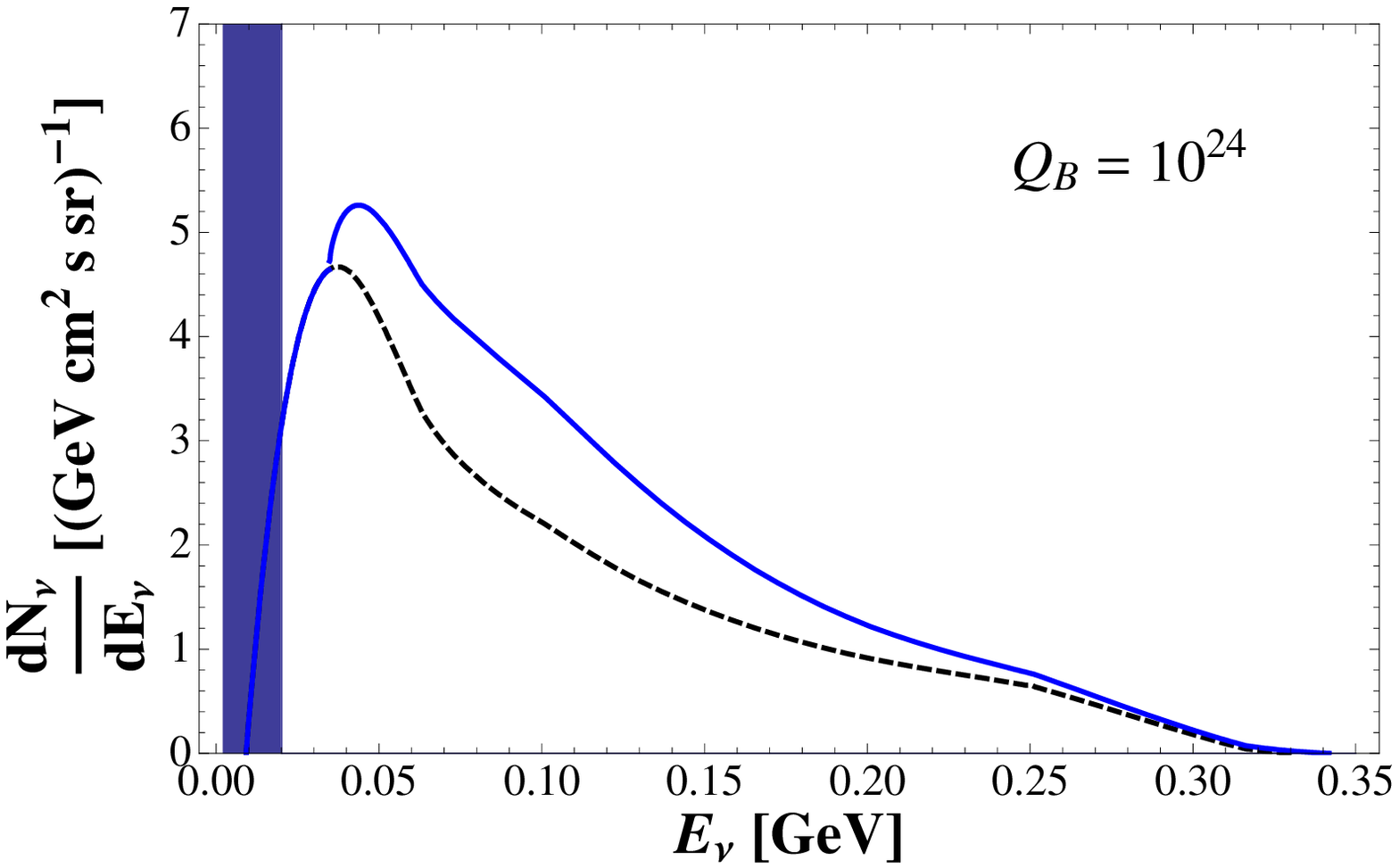,width=1.0\linewidth,clip=} \\
\epsfig{file=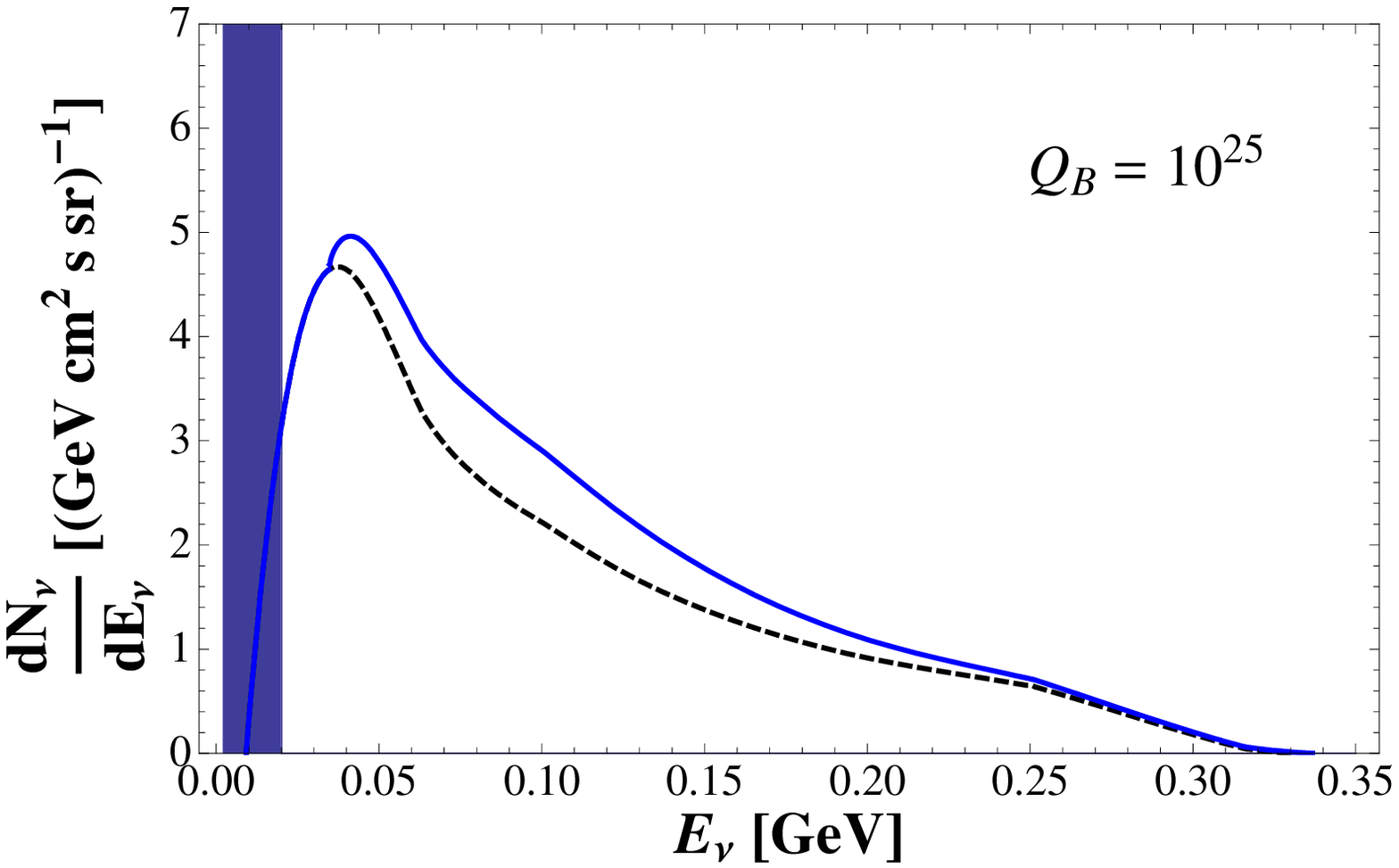,width=1.0\linewidth,clip=}
\end{tabular}
\caption{Neutrino flux spectra averaged over zenith angles, in the presence of relic Q-balls (the upper line) and in the absence of relic Q-balls (the lower line), for two values of the baryon number, $Q_B=10^{24}$ and $Q_B=10^{25}$.  The atmospheric neutrino flux shown by the lower curve is based on Ref.~\cite{Honda:1995hz}.  
}
\label{fig1}
\end{figure}

The atmospheric neutrino flux has a peak at $0.1\, {\rm GeV}$ reaching $\sim 1\, {\rm cm}^{-2} {\rm s}^{-1} $~\cite{PhysRevD.37.122,Barr:1989ru,Honda:1995hz,Barr:2004br,Bergstrom:1999kd}.  The flux of solar neutrinos dominates at  low energies, below $10~{\rm MeV}$, as shown in Fig.~\ref{fig1} by a dark vertical band.  The flux from Q-balls gives a non-negligible contribution at energies $ 0.01-0.3~{\rm GeV}$. 

Since the Q-ball interaction rate is proportional to the density, 
this flux is enhanced by the higher density in the core.  For simplicity we treat the inner and outer cores as one, and we approximate the Earth density profile by a single step function:  
\be \rho_{\earth} (r)= \left\{
     \begin{array}{lr}
       \rho_{\rm c} = 12~ {\rm g}/ {\rm cm}^{3},  &  r \leq R_{\rm c} \\
       \rho_{\rm m} = 3~{\rm g}/ {\rm cm}^{3},  &  R_{\rm c} \leq r \leq R_{\earth}, 
     \end{array}
   \right.
   \label{rho}
\ee
where $R_{\rm c} = 3.4 \times 10^{3} {\rm km}$ and $R_{\rm \earth} = 6.4 \times 10^{3} {\rm km}$ .   

The neutrinos are produced in $ p \bar{p}$ and $ n \bar{n}$ annihilations, which, on average generate $\langle n_{\pi} \rangle \approx 5$ pions per nucleon \cite{Blumel:1994yj}. The subsequent decays of the pions produce neutrinos with energies $E_{\nu} \sim 0.1~{\rm GeV}$.  We will see that this simple estimate is born out in the more detailed calculation of the neutrino spectrum to follow. 

The primary background to the Q-ball induced neutrino effect is the atmospheric background. Extensive Monte Carlo work has been devoted to the subject of calculating this flux, as it is important is being able to observe neutrino oscillations. We use the fluxes from  Refs.~\cite{PhysRevD.37.122,Barr:1989ru,Honda:1995hz,Barr:2004br}. The inclusion of geomagnetic effects is important in a more detailed analysis.

The neutrino spectrum from Q-ball catalyzed antiprotons will depend on the pion spectrum from $p \bar{p}$ annihilations. For simplicity we focus on the sum over neutrino flavors, ignoring the neutrino oscillations:
\be \frac{dN_{\nu}}{dE_{\nu}} =  \frac{\partial N_{\nu}}{\partial N_{\pi}} \frac{\partial k_{\pi}}{\partial E_{\nu}} \left( \frac{dN_{\pi}}{dk_{\pi}} \right), \ee
where the pion spectrum is found from the data on $p\bar{p}$ annihilation at rest~\cite{Dover:1992vj}: 

\be \frac{dN_{\pi}}{dk_{\pi}} = c k^{2} e^{-\alpha \omega} .\ee
Here $\omega \equiv \sqrt{m_{\pi}^{2} + k^{2}}$, and $\alpha \approx 7 ~ {\rm GeV}^{-1}$, and $c\sim 90~{\rm GeV}^{-3}$ normalizes the spectrum so that $N_{\pi} \sim 5$. All momenta and energies are measured in the laboratory frame. In the cascade $\pi^{\pm} \rightarrow \mu^{\pm} \nu_{\mu}$,  $\mu \rightarrow e + \nu_{e} + \nu_{\mu}$ , the pion energy is split approximately uniformly among the four final decay products, so we have $ E_{\nu} = \frac{1}{4} \sqrt{k^2 + m_{\pi}^{2}}$.  In Fig.~\ref{fig1} we show the expected spectrum $dN_{\nu}/d E_{\nu}$ normalzed to give the correct total flux from Q-balls, $F_Q= \int (dN_{\nu}/d E_{\nu})dE_\nu$.    
 To find the evidence of Q-balls, one must be able to discriminate between the ordinary atmospheric spectrum $F_{atm}$ and $F_{\rm tot} \equiv F_{\rm atm} + F_{Q}$.  The Q-ball induced flux peaks at $E_{\nu} \sim 80~ {\rm MeV}$.

\begin{figure}[t]
\centering
\epsfig{file=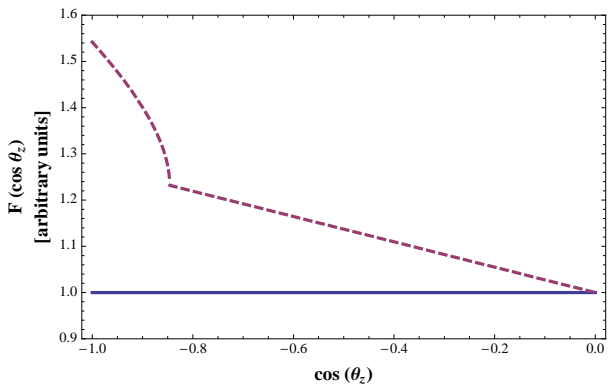, width=8cm}
\caption{Zenith angle dependence of the neutrino flux due to Q-balls, summed over neutrino flavors. The enhanced flux close to nadir is due to the increased matter density near the Earth's core.  For individual neutrino flavors, neutrino oscillations would affect the zenith angle dependence. }
\label{fig2}
\end{figure}

The isotropy of the Q-ball flux on the Earth implies that the zenith angle dependence of $F_{Q}$ is determined entirely by the Earth geometry and density distribution: 
\be 
   F_{\nu}(\theta_{z}) \propto \left\{
     \begin{array}{l}
 0 , \ \ \ \theta_{z} < \pi/2 \\
         \rho_{\rm m} \cos \theta_{z} ,  \ \ \  \pi/2 \leq \theta_{z} \leq \theta_{\rm c} \\
          \rho_{\rm m}  \cos \theta_{z}  
        + ( \rho_{\rm c} -\rho_{\rm m} ) f_{c} ( \theta_{z}) ,  \  \theta_{\rm c} \leq \theta_{z} \leq \pi ,
     \end{array}
   \right.
   \label{zen}
\ee
where $f_{c} (\theta_{z})= \sqrt{(R_{\rm c}/R_{\earth})^2 - \sin^{2} \theta_{z}}$,  and $\theta_{\rm c} \equiv \arcsin (R_{\rm c}/R_{\earth})=32^\circ $ is defined as the nadir angle that grazes the core. We show this zenith angle dependence in Fig.~\ref{fig2}. 

The additional upgoing neutrino flux due to Q-balls changes the downward-upward asymmetry observed by Super-Kamiokande and other experiments, which is usually characterized in terms of $R= N_{up}/N_{down}$ and $A = \frac{N_{up}-N_{down}}{N_{up}+N_{down}}$, where upward (downward) going events are those with $-1 < \cos \theta_{z} < -0.2$ ($0.2 < \cos \theta_{z} < 1$). 
Though we have ignored neutrino oscillations and geomagnetic effects, both of these introduce additional zenith angle dependence. For example, Monte Carlo simulations \cite{Barr:2004br,Giunti:2007ry} indicate that Soudan should see an asymmetry $A_{\mu} = - 0.4$ at $E_{\nu} = 0.1~ {\rm GeV}$, whereas Super-Kamiokande sees $A_{\mu} = +0.4$ at the same energy.  Given equal other conditions,  it is probably easier to observe the neutrino flux from dark matter Q-balls in a location where the geomagnetic effects suppress the upward going neutrinos relative to downward going neutrinos.

Annual modulation of the dark-matter particle flux due to the Earth motion implies a corresponding modulation of the neutrino flux from Q-balls of the order of 10\%.

To summarize, we propose a new experimental probe on dark matter in the form of supersymmetric Q-balls. The relic Q-balls passing through Earth convert nucleons into antinucleons.  The $p\bar{p}$ and  $n\bar{n}$ annihilations in the wake of the passing Q-balls generate a neutrino flux with a peculiar zenith angle dependence and a small annual modulation.  A detailed analysis of the present and future neutrino data can set limits on the relic Q-balls, or it can lead to a discovery of this form of supersymmetric dark matter. 

The authors thank John M. Cornwall for helpful discussions. 
This work was supported in part  by DOE grant DE-FG03-91ER40662 and by the NASA ATFP grant  NNX08AL48G.

\bibliography{qball}
\bibliographystyle{h-physrev3}

\end{document}